\documentstyle[preprint,aps]{revtex}

\begin{document}

\draft

\title{Mirror matter admixtures in $\Omega^-$ two body non-leptonic decays and
the $|\Delta I|=1/2$ rule}

\author{
G.~S\'anchez-Col\'on\cite{email1},
R.~Huerta
}
\address{
Departamento de F\'{\i}sica Aplicada.\\
Centro de Investigaci\'on y de Estudios Avanzados del IPN. Unidad
M\'erida.\\
A.P. 73, Cordemex. M\'erida, Yucat\'an, 97310. MEXICO.
}

\author{
A.~Garc\'{\i}a
}
\address{
Departamento de F\'{\i}sica.\\
Centro de Investigaci\'on y de Estudios Avanzados del IPN.\\
A.P. 14-740. M\'exico, D.F., 07000. MEXICO.
}

\date{\today}

\maketitle

\begin{abstract}
We  extend our previous analysis of mirror matter admixtures to reproduce
the $|\Delta I|=1/2$ rule predictions in the decays $\Omega^-\to\Xi^-\pi^0$,
$\Omega^-\to\Xi^0\pi^-$, and $\Omega^-\to\Lambda K^-$. The results are
satisfactory, lending credibility to the possibility that the enhancement
phenomenon may be attributed to such admixtures.
\end{abstract}


One of the main current interests is the detection of new physics. Its
existence and its connection with the physics of the Standard Model may help
solve some of the long standing problems in low energy physics, such as the
enhancement phenomenon observed in meson and hyperon decays~\cite{okun}.
However, the success of the standard model indicates that new forms of
matter, if any, are far away and they will be difficult to discover.
Nevertheless, this does not exclude the possibility that such new physics
manifests itself indirectly in our low energy world. A great oportunity is
provided by the enormous gap that exists from strong and electromagnetic
interactions to weak interactions. Even a small admixture between ordinary and
new matter may show its presence somewhere within this gap. It could lead to
effects beyond the standard model predictions.

One particularly interesting possibility, already discussed well before the
advent of the standard model by Lee and Yang~\cite{lee} in their pioneering
paper on parity violation, is the existence of mirror matter. This possibility
has been discussed systematically eversince~\cite{mohapatra,barr}. In recent
papers we have studied in detail the effects of mirror hadrons in non-leptonic
and weak radiative decays of hyperons~\cite{wnlm,apriori,9945,L1} and
non-leptonic decays of pseudoscalar mesons~\cite{mpla001749}. Indeed,
tiny admixtures with ordinary hadrons may easily produce observable
effects in such decays. This led us to establish a lower bound on the
mass of a mirror proton of about $10^6{\rm GeV}$~\cite{bound}. Also,
as mentioned above, one may demand of new physics to help us solve
long standing problems. One such problem is the enhancement
accompanying the $|\Delta I|=1/2$ rule observed in many non-leptonic
decays. We have shown using an  ansatz that, indeed, this rule can be
reproduced. If this new effect in low energy physics due to such
admixtures is to provide an explanation of the $|\Delta I|=1/2$ rule
one must expect that the formulation discussed previously can be
extended to cover the two body non-leptonic decays of $\Omega^-$.

It is the purpose of this letter to extend our previous analysis to
$\Omega^-$ decays. We shall show that the $|\Delta I|=1/2$ rule for
these non-leptonic decays can be reproduced by such small admixtures.
These latter we refered to as {\it a priori }mixings.

\paragraph*{A priori mixings in $\Omega^-$.}
To implement the mixing between ordinary and mirror hadrons we have used an
ansatz, described in Ref.~\cite{apriori}. Here we shall review it briefly, more
details may be found there. Mirror hadrons are assumed to have parities opposite
to the ones of their ordinary matter counterparts. There exists a connection
between the two worlds which leads to an off-diagonal piece in the
mass operator. The final diagonalization of this operator is performed by
rotation operators $R$ and $R^M$ that yield small admixtures of mirror hadrons
in the physical (mass eigenstates) ordinary hadrons. These operators are assumed
to obey $U$-spin type ladder transformation properties and both the ordinary
spin 3/2 resonances and their mirror counterparts belong to decuplet $SU(3)$
representations (or to 20-plets $SU(4)$ representations which include charm).
Skipping details, the action of the several $U$-type $R$ and $R^M$ matrices upon
these representations leads to a physical $\Omega^-$ with mirror matter
admixtures, namely,

\begin{equation}
\Omega^-_{ph} =
\Omega^-_{s} +
\frac{a}{2}g_{10}\Xi^{*-}_{s} +
\frac{c'}{2}g^0_{10}\Xi^{*-}_{p} +
\cdots.
\label{omegap}
\end{equation}

\noindent
The coefficients $a$ and $c'$ were defined in Ref.~\cite{apriori}, they belong
to the expansion of $R$ and $R^M$ in terms of $U$-spin type operators. Here
$g_{10}$ is a reduced matrix element between states in the same 10
representation and $g^0_{10}$ is a reduced matrix element between states in this
10 and the mirror 10 representations. In order to make connection with our
previous work we must identify

\begin{equation}
\sigma = a(-\frac{1}{2\sqrt{3}}g_{10}),
\label{sigma}
\end{equation}

\begin{equation}
\delta' = c'(\frac{1}{2\sqrt{3}}g^0_{10}),
\label{delta'}
\end{equation}

\noindent
and then (\ref{omegap}) becomes

\begin{equation}
\Omega^-_{ph} =
\Omega^-_{s} -
\sqrt{3}\sigma\Xi^{*-}_{s} +
\sqrt{3}\delta'\Xi^{*-}_{p} +
\cdots.
\label{omega}
\end{equation}

\noindent
The subindeces $s$ and $p$ stand for positive and negative intrinsic parities,
respectively,  and the dots in (\ref{omegap}) and (\ref{omega}) refer to other
mixings (with strong flavors other than strangeness) which are not relevant
here. Our phase conventions are those of Ref.~\cite{gibson}.

We must point out that the identifications (\ref{sigma}) and (\ref{delta'}) have
a parallelism at the quark level~\cite{rmf45}. This means that they are
necessary if one hopes to ever develop a formulation of mirror matter admixtures
starting at this level~\cite{barr}.

\paragraph*{Two-body non-leptonic decays amplitudes of $\Omega^-$.}
We shall study the decays $\Omega^-\to\Xi^-\pi^0$,
$\Omega^-\to\Xi^0\pi^-$, and $\Omega^-\to\Lambda K^-$. They are described by a
Lorentz invariant amplitude of the form

\begin{equation}
\langle M(q) B'(p')  |H| B(p) \rangle =
\bar{u}(p') ( {\cal B} + \gamma^5 {\cal C} ) q^{\mu} u_{\mu}(p),
\label{amplitude}
\end{equation}

\noindent
where $u$ and $u_{\mu}$ are Dirac and Rarita-Schwinger spinors, respectively,
and ${\cal B}$ and ${\cal C}$ are $p$-wave (parity conserving) and $d$-wave
(parity violating) amplitudes, respectively. $H$ is the transition operator and
$B(p)$, $B'(p')$, and $M(q)$ represent $s=3/2$, $s=1/2$ baryons, and a
pseudoscalar meson, respectively.

In our approach the transition operator $H$ is the strong flavor and parity
conserving interaction hamiltonian responsible for the two-body strong decays of
the other $s=3/2$ resonances in the decuplet where $\Omega^-_s$ belongs to. In
this letter we shall neglect isospin breaking, i.e. we shall assume
$H$ is a $SU(2)$ scalar.

In order to obtain explicit expressions for ${\cal B}$ and ${\cal C}$ we need
the mirror admixtures in $\Xi^-$, $\Xi^0$, $\Lambda$, $\pi^0$, $\pi^-$, and
$K^-$. We take them from our previous work~\cite{apriori}, namely

\begin{equation}
\Xi^-_{ph} =
\Xi^-_s -
\sigma\Sigma^-_s +
\delta'\Sigma^-_p
+ \cdots
\label{xi-}
\end{equation}

\begin{equation}
\Xi^0_{ph} =
\Xi^0_s -
\sigma
(\frac{1}{\sqrt 2}\Sigma^0_s + \sqrt{\frac{3}{2}}\Lambda_s) +
\delta'
(\frac{1}{\sqrt 2}\Sigma^0_p + \sqrt{\frac{3}{2}}\Lambda_p)
+ \cdots
\label{xi0}
\end{equation}

\begin{equation}
\Lambda_{ph} =
\Lambda_s +
\sigma\sqrt{\frac{3}{2}}(\Xi^0_s - n_s) +
\delta\sqrt{\frac{3}{2}}\Xi^0_p +
\delta'\sqrt{\frac{3}{2}}n_p
+ \cdots
\label{lambda}
\end{equation}

\begin{equation}
\pi^0_{ph} =
\pi^0_{p} -
\sigma\frac{1}{\sqrt 2}( K^0_{p} + \bar K^0_{p} ) +
\delta\frac{1}{\sqrt 2}( K^0_{s}- \bar K^0_{s} ) + \cdots
\label{pi0}
\end{equation}

\begin{equation}
\pi^-_{ph} =
\pi^-_{p} + \sigma K^-_{p} + \delta K^-_{s} + \cdots
\label{pi-}
\end{equation}

\begin{equation}
K^-_{ph} =
K^-_p -
\sigma \pi^-_p +
\delta' \pi^-_s +
\cdots
\label{k-}
\end{equation}

\noindent
The states (\ref{omega}), (\ref{xi-}) -- (\ref{k-}), and the properties of $H$
discussed above, lead to

\begin{equation}
{\cal B}(\Omega^-\to\Xi^-\pi^0) =
-\sigma
(
 \sqrt 3 g_{{}_{\pi^0_p\Xi^-_s,\Xi^{*-}_s}} +
\frac{1}{\sqrt{2}}g_{{}_{\bar{K}^0_p\Xi^-_s,\Omega^-_s}}
) ,
\label{bxi-}
\end{equation}

\begin{equation}
{\cal C}(\Omega^-\to\Xi^-\pi^0) =
\delta'
\sqrt{3} g_{{}_{\pi^0_p\Xi^-_s,\Xi^{*-}_p}} -
\delta\frac{1}{\sqrt{2}}g_{{}_{\bar{K}^0_s\Xi^-_s,\Omega^-_s}}
,
\label{cxi-}
\end{equation}

\begin{equation}
{\cal B}(\Omega^-\to\Xi^0\pi^-) =
\sigma
(
- \sqrt 3 g_{{}_{\pi^-_p\Xi^0_s,\Xi^{*-}_s}} +
g_{{}_{K^-_p\Xi^0_s,\Omega^-_s}}
) ,
\label{bxi0}
\end{equation}

\begin{equation}
{\cal C}(\Omega^-\to\Xi^0\pi^-) =
\delta'
\sqrt{3} g_{{}_{\pi^-_p\Xi^0_s,\Xi^{*-}_p}} +
\delta g_{{}_{K^-_s\Xi^0_s,\Omega^-_s}}
,
\label{cxi0}
\end{equation}

\begin{equation}
{\cal B}(\Omega^-\to\Lambda K^-) =
\sigma
(
- \sqrt 3 g_{{}_{K^-_p\Lambda_s,\Xi^{*-}_s}} +
\sqrt{\frac{3}{2}}g_{{}_{K^-_p\Xi^0_s\Omega^-_s}}
) ,
\label{bl}
\end{equation}

\begin{equation}
{\cal C}(\Omega^-\to\Lambda K^-) =
\delta'
\sqrt{3} g_{{}_{K^-_p\Lambda_s,\Xi^{*-}_p}} +
\delta
\sqrt{\frac{3}{2}}g_{{}_{K^-_p\Xi^0_p,\Omega^-_s}}
.
\label{cl}
\end{equation}

\noindent
The constants $g_{{}_{M_pB'_s,B_s}}$ are the Yukawa strong couplings observed in
the strong two-body decays of $s=3/2$ resonances. The constants
$g_{{}_{M_pB'_s,B_p}}$,  $g_{{}_{M_sB'_s,B_s}}$ and $g_{{}_{M_pB'_p,B_s}}$ are
new, because they involve mirror matter. However, because of our assumptions
they all have common properties. In the isospin limit they are related by

\begin{equation}
g_{{}_{K^-\Xi^0,\Omega^-}}=g_{{}_{\bar{K}^0\Xi^-,\Omega^-}},
\label{su2a}
\end{equation}

\begin{equation}
g_{{}_{\pi^-\Xi^0,\Xi^{*-}}}=-\sqrt{2}g_{{}_{\pi^0\Xi^-,\Xi^{*-}}},
\label{su2b}
\end{equation}

\noindent
where the subindeces $s$ and $p$ may be dropped out. In the $SU(3)$ limit one
also has

\begin{equation}
g_{{}_{\pi^0\Xi^-,\Xi^{*-}}} = -\frac{1}{\sqrt{6}}
g_{{}_{\bar{K}^0\Xi^-,\Omega^-}},
\label{su3a}
\end{equation}

\begin{equation}
g_{{}_{K^-\Lambda,\Xi^{*-}}} =
\frac{1}{\sqrt{2}} g_{{}_{\bar{K}^0\Xi^-,\Omega^-}}.
\label{su3b}
\end{equation}

\paragraph*{The $|\Delta I|=1/2$ rule and other properties of $\Omega^-$ decay
amplitudes.}
In the explicit amplitudes (\ref{bxi-}) -- (\ref{cl}) we can now replace the
isospin equalities (\ref{su2a}) and (\ref{su2b}), with the $s$ and $p$
subindeces restored. Eqs.~(\ref{bxi-}) -- (\ref{cl}) become

\begin{equation}
{\cal B}(\Omega^-\to\Xi^-\pi^0) =
-\sigma
(
 \sqrt 3 g_{{}_{\pi^0\Xi^-,\Xi^{*-}}} +
\frac{1}{\sqrt{2}}g_{{}_{\bar{K}^0\Xi^-,\Omega^-}}
) ,
\label{bxi-2}
\end{equation}

\begin{equation}
{\cal C}(\Omega^-\to\Xi^-\pi^0) =
\delta'
\sqrt{3} g_{{}_{\pi^0_p\Xi^-_s,\Xi^{*-}_p}} -
\delta
\frac{1}{\sqrt{2}}g_{{}_{\bar{K}^0_s\Xi^-_s,\Omega^-_s}}
,
\label{cxi-2}
\end{equation}

\begin{equation}
{\cal B}(\Omega^-\to\Xi^0\pi^-) =
\sigma
(
\sqrt{6}  g_{{}_{\pi^0\Xi^-,\Xi^{*-}}} +
g_{{}_{\bar{K}^0\Xi^-,\Omega^-}}
) ,
\label{bxi02}
\end{equation}

\begin{equation}
{\cal C}(\Omega^-\to\Xi^0\pi^-) =
- \delta'
\sqrt{6} g_{{}_{\pi^0_p\Xi^-_s,\Xi^{*-}_p}} +
\delta
g_{{}_{\bar{K}^0_s\Xi^-_s,\Omega^-_s}}
,
\label{cxi02}
\end{equation}

\begin{equation}
{\cal B}(\Omega^-\to\Lambda K^-) =
\sigma
(
- \sqrt 3 g_{{}_{K^-\Lambda,\Xi^{*-}}} +
\sqrt{\frac{3}{2}}g_{{}_{\bar{K}^0\Xi^-\Omega^-}}
) ,
\label{bl2}
\end{equation}

\begin{equation}
{\cal C}(\Omega^-\to\Lambda K^-) =
\delta'
\sqrt{3} g_{{}_{K^-_p\Lambda_s,\Xi^{*-}_p}} +
\delta
\sqrt{\frac{3}{2}}g_{{}_{\bar{K}^0_p\Xi^-_p,\Omega^-_s}}
.
\label{cl2}
\end{equation}

\noindent
We have omitted the subindeces $s$ and $p$ in the $g$'s of the ${\cal B}$
amplitudes because the states involved carry the normal intrinsic parities of
hadrons.

One readily sees that the equalities

\begin{equation}
\frac{{\cal B}(\Omega^-\to\Xi^-\pi^0) }{{\cal B}(\Omega^-\to\Xi^0\pi^-)} =
\frac{{\cal C}(\Omega^-\to\Xi^-\pi^0) }{{\cal C}(\Omega^-\to\Xi^0\pi^-)} =
-\frac{1}{\sqrt{2}}
\label{deltai}
\end{equation}

\noindent
are obtained. This is the so called $|\Delta I|=1/2$ rule in these decay modes
and applies to both decay amplitudes. The mirror admixtures in $\Omega^-$ lead
to the results of this rule. However, it must be recalled that in this approach
these results are really a $\Delta I=0$ rule, since isospin is assumed unbroken.

If we now assume the $SU(3)$ limit, we can replace (\ref{su3a}) and (\ref{su3b})
in the amplitudes (\ref{bxi-2}) -- (\ref{cl2}). In this case one obtains

\begin{equation}
{\cal B}(\Omega^-\to\Xi^-\pi^0) =
{\cal B}(\Omega^-\to\Xi^0\pi^-) = {\cal B}(\Omega^-\to\Lambda K^-) = 0,
\label{su3lmt1}
\end{equation}

\begin{equation}
{\cal C}(\Omega^-\to\Xi^-\pi^0) =
- \frac{1}{\sqrt{2}} (\delta' g_{{}_{\bar{K}^0_p\Xi^-_s,\Omega^-_p}}
+ \delta g_{{}_{\bar{K}^0_s\Xi^-_s,\Omega^-_s}}),
\label{su3lmt2}
\end{equation}

\begin{equation}
{\cal C}(\Omega^-\to\Xi^0\pi^-) =
\delta' g_{{}_{\bar{K}^0_p\Xi^-_s,\Omega^-_p}}
+ \delta g_{{}_{\bar{K}^0_s\Xi^-_s,\Omega^-_s}},
\label{su3lmt3}
\end{equation}

\begin{equation}
{\cal C}(\Omega^-\to\Lambda K^-) =
\sqrt{\frac{3}{2}} (\delta' g_{{}_{\bar{K}^0_p\Xi^-_s,\Omega^-_p}}
+ \delta g_{{}_{\bar{K}^0_p\Xi^-_p,\Omega^-_s}}).
\label{su3lmt4}
\end{equation}

\noindent
In this limit we see that the parity-conserving amplitudes vanish while the
parity-violating ones remain non-zero. This is the same result obtained as a
theorem previously~\cite{limit}. It should be contrasted with the theorem of
Hara~\cite{hara}, valid in $W$-mediated non-leptonic decays. This theorem
states that in this limit it is the parity-violating amplitudes that vanish and
the parity-conserving ones that remain non-zero.

The above formulation with mirror admixtures in $\Omega^-$ shows that our
earlier result for non-leptonic and weak radiative decays of
hyperons~\cite{wnlm,apriori,9945,L1} and non-leptonic decays of
pseudoscalar mesons~\cite{mpla001749} are equally applicable in
$\Omega^-$ non-leptonic decays.

\paragraph*{Comparison with experiment.}
So far our results are analytic. For completeness, it is necessary
to verify that experimental results can be reproduced with reasonable values of
the several parameters that appear in amplitudes (\ref{bxi-}) -- (\ref{cl}). By
this we mean that (i) the {\it a priori }mixing angles $\sigma$, $\delta$, and
$\delta'$ must be constrained to the values obtained before and that (ii) the
Yukawa coupling constants have values comparably close to the already observed
ones.

The observables in the three decays we discuss are the partial decay rates and
the asymmetries. Their measured values are displayed in Table~\ref{table1}. In
terms of the decay amplitudes these observables are given by $\Gamma =
(q^3/12\pi M) [(E+m) |{\cal B}|^2 + (E-m) |{\cal C}|^2]$ and $\alpha = (2|{\cal
B}||c\,{\cal C}|)/(|{\cal B}|^2 + |c\,{\cal C}|^2)$. Here, $q^2 = (E-m)(E+m)$,
$E\pm m = [(M \pm m)^2 - m^2_o]/2M$, $c^2=(E - m)/(E + m)$, and $M$, $m$,
and $m_o$ are the masses of the $s=3/2$, $s=1/2$ baryons and the
pseudoscalar meson, respectively. In addition we should take into
acount the experimental measurement (also displayed in
Table~\ref{table1}) of the combined decay rate
$\Gamma(\Xi^{*-}\to\Xi\pi) = \Gamma(\Xi^{*-} \to \Xi^-\pi^0) +
\Gamma(\Xi^{*-} \to \Xi^0\pi^-)$, which constrains the couplings
$g_{{}_{\pi^0\Xi^-,\Xi^{*-}}}$ and $g_{{}_{\pi^-\Xi^0,\Xi^{*-}}}$.
These rates are given by the above formula when one inserts ${\cal B}
= g_{{}_{\pi^0\Xi^-,\Xi^{*-}}},\ g_{{}_{\pi^-\Xi^0,\Xi^{*-}}}$ and
${\cal C} = 0$. The restrictions on the {\it a priori }mixing angles
from hyperon weak radiative and non-leptonic decays
are~\cite{9945,L1} $\sigma = (4.9\pm 2.0) \times 10^{-6}$, $\delta =
(0.22\pm 0.09)\times 10^{-6}$, and $\delta' = (0.26\pm 0.09)\times
10^{-6}$. We shall make a $\chi^2$ fit. We have 10 restrictions to
form the $\chi^2$ function, the 7 experimental measurements and these
three restrictions on the angles.

Our interest here is to explore the possible success of the above formulation of
the $|\Delta I|=1/2$ rule. Accordingly, we shall work in the $SU(2)$  symmetry
limit and use Eqs.~(\ref{bxi-2}) -- (\ref{cl2}). These amplitudes depend on 10
parameters, 3 ordinary Yukawa couplings ($g_{{}_{\pi^0\Xi^-,\Xi^{*-}}}$,
$g_{{}_{\bar{K}^0\Xi^-,\Omega^-}}$, and $g_{{}_{K^-\Lambda,\Xi^{*-}}}$), 4
mirror couplings involving mirror hadrons ($g_{{}_{\pi^0_p\Xi^-_s,\Xi^{*-}_p}}$,
$g_{{}_{\bar{K}^0_s\Xi^-_s,\Omega^-_s}}$, $g_{{}_{K^-_p\Lambda_s,\Xi^{*-}_p}}$,
and $g_{{}_{\bar{K}^0_p\Xi^-_p,\Omega^-_s}}$) and the angles $\sigma_\Omega$,
$\delta_\Omega$, and $\delta'_\Omega$. To avoid confusions we add the subindex
$\Omega$ when the angles are used as free parameters through Eqs.~(\ref{bxi-2})
-- (\ref{cl2}). We shall deal with the mirror couplings following the analysis
of Ref.~\cite{9945}. As was discussed there, the assumption that strong
interactions are the same for ordinary and mirror matter allows us to put the
magnitudes of the mirror couplings equal to their counterparts involving
ordinary hadrons. However, their signs may differ. We must allow two signs in
front of each of the magnitudes of the four mirror couplings. This way we have 6
free parameters, the 3 ordinary couplings and the 3 mixing angles, and the signs
of the mirror couplings. Our $\chi^2$ fit consists of fitting these 6
parameters to a $\chi^2$ with 10 constraints, allowing for each possible choice
for the relative signs between ordinary and mirror couplings. There is no
essential difficulty in doing this, although it is a tedious task.

Our best result yields the predictions displayed in Table~\ref{table1}. The
total $\chi^2$ is 25.16. The 6 parameters take the values,
$g_{{}_{\pi^0\Xi^-,\Xi^{*-}}} = 4.326\pm 0.401$,
$g_{{}_{\bar{K}^0\Xi^-,\Omega^-}} = -10.35\pm 1.01$,
$g_{{}_{K^-\Lambda,\Xi^{*-}}} = -7.773\pm 0.592$, $\sigma_\Omega =
(5.1\pm 2.8)\times 10^{-6}$, $\delta_\Omega = (0.263\pm 0.067)\times
10^{-6}$, and $\delta'_\Omega = (0.215\pm 0.064)\times 10^{-6}$. These
couplings are in ${\rm GeV}^{-1}$. The signs of the mirror couplings
of this fit are given by the equalities
$g_{{}_{\pi^0_p\Xi^-_s,\Xi^{*-}_p}} = g_{{}_{\pi^0\Xi^-,\Xi^{*-}}}$,
$g_{{}_{\bar{K}^0_s\Xi^-_s,\Omega^-_s}} = -
g_{{}_{\bar{K}^0_p\Xi^-_p,\Omega^-_s}} = -
g_{{}_{\bar{K}^0\Xi^-,\Omega^-}}$, and
$g_{{}_{K^-_p\Lambda_s,\Xi^{*-}_p}} = -
g_{{}_{K^-\Lambda,\Xi^{*-}}}$.

\paragraph*{Discussion.}
In this letter we have studied how to extend our previous approach to
non-leptonic and weak radiative decays to the two-body non-leptonic
decays of $\Omega^-$. This extension is necessary if one expects that the
enhancement phenomenon observed in these decays may be attributed to small
admixtures of mirror hadrons with ordinary ones.

Assuming the $SU(2)$ symmetry limit, the predictions of the $|\Delta I|=1/2$
rule accompanying this enhancement are reproduced for both the parity
conserving and parity violating amplitudes. In the $SU(3)$ limit the former
amplitudes vanish, as was the case in our earlier work. The fitted values of
$\sigma_\Omega$, $\delta_\Omega$, and $\delta'_\Omega$ are consistent with the
previous values of the mixing angles which we use as constraints, within the
errors bars of the latter. This points towards a universality property of these
angles, a feature which is necessary if one is ever to be able to obtain these
admixtures starting at the quark level. The experimental decay rates of
$\Omega^- \to \Xi^- \pi^0$ and $\Omega^-\to\Xi^0\pi^-$ are reproduced within
$20\%$ and $15\%$, respectively. While the other observables are reproduced
within a very small percentage. The three Yukawa couplings take values that are
of the correct order of magnitude, compared to reported Yukawa couplings of
$\Delta^{++} \to p \pi^+$ of $15.7{\rm GeV}^{-1}$~\cite{lopez}. Also, these
values are consistent with small breaking of the  $SU(3)$ symmetry
relationships (\ref{su3a}) and (\ref{su3b}). The results are satisfactory.

We have limited our analysis to the validity of the $|\Delta I|=1/2$ rule in the
enhancement phenomenon and not to its breaking. In $\Omega^-$ decays this rule
is experimentally observed to be valid within a $35\%$ in the quotient of branching
ratios of $\Omega^- \to \Xi^- \pi^0$ and $\Omega^-\to\Xi^0\pi^-$,
which here is reflected in the above $20\%$ and $15\%$ deviations (or the
corresponding $\Delta\chi^2=14.67$ and $\Delta\chi^2=9.96$). In the near future,
in a separate cover we shall address these deviations.

The authors are grateful to CONACyT (M\'exico) for partial support.

\begin{table}
\caption{Experimentally measured, predicted values and
$\Delta\chi^2$ contributions of the observables in the decays
$\Omega^- \to \Xi^-\pi^0$, $\Omega^- \to \Xi^0\pi^-$, $\Omega^- \to
\Lambda K^-$, and $\Xi^{*-}\to\Xi\pi$, indicated by the subindeces 1,
2, 3, and 4, respectively.}
\label{table1}
\begin{tabular}{l c d d}
\hline
Decay & Experiment & Prediction & $\Delta\chi^2$ \\
\hline
$\Gamma_1(10^9{\rm sec}^{-1})$ & $1.046\pm 0.051$ [15] & 1.241 &
14.64\\
$\alpha_1$ & $0.05\pm 0.21$ [15] & 0.076 & 0.02 \\
$\Gamma_2(10^9{\rm sec}^{-1})$ & $2.871\pm 0.095$ [15] & 2.571 &
9.96 \\
$\alpha_2$ & $0.09\pm 0.14$ [15] & 0.078 & 0.01 \\
$\Gamma_3(10^9{\rm sec}^{-1})$ & $8.25\pm 0.15$ [15] & 8.247 &
0.0004 \\
$\alpha_3$ & $-$$0.026\pm 0.026$ [15--17] & $-$0.021 & 0.04 \\
$\Gamma_4({\rm MeV})$ & $9.9\pm 1.9$ [18--21] & 9.8 & 0.001 \\
\hline
\end{tabular}
\end{table}


\begin{references}

\bibitem[*]{email1}
e-mail address: gsanchez@mda.cinvestav.mx

\bibitem{okun}
For a review and an extended list of references see:
L.~B.~Okun,
{\it Leptons and quarks\ }
(Elsevier Science, New York, 1982).

\bibitem{lee}
T.~D.~Lee, and C.~N.~Yang,
Phys. Rev. {\bf 104} (1956) 254.

\bibitem{mohapatra}
R.~N.~Mohapatra,
{\it Unification and Supersymmetry. The frontiers of quark-lepton physics,}
Contemporary Physics
(Springer-Verlag, N.~Y., 1986).

\bibitem{barr}
S.~M.~Barr, D.~Chang, and G.~Senjanovi\'c,
Phys.\ Rev.\ Lett.\ {\bf 67}\ (1991)\ 2765.

\bibitem{wnlm}
A.~Garc\'{\i}a, R.~Huerta, and G.~S\'anchez-Col\'on,
Rev.\ Mex.\ Fis.\ {\bf 43}\ (1997)\ 232.

\bibitem{apriori}
A.~Garc\'{\i}a, R.~Huerta, and G.~S\'anchez-Col\'on,
J.\ Phys.\ G:\ Nucl.\ Part.\ Phys.\ {\bf 24}\ (1998)\ 1207.

\bibitem{9945}
A.~Garc\'{\i}a, R.~Huerta, and G.~S\'anchez-Col\'on,
J.\ Phys.\ G:\ Nucl.\ Part.\ Phys.\ {\bf 25}\ (1999)\ 45.

\bibitem{L1}
A.~Garc\'{\i}a, R.~Huerta, and G.~S\'anchez-Col\'on,
J.\ Phys.\ G:\ Nucl.\ Part.\ Phys.\ {\bf 25}\ (1999)\ L1.

\bibitem{mpla001749}
A.~Garc\'{\i}a, R.~Huerta, and G.~S\'anchez-Col\'on,
Mod.\ Phys.\ Lett.\ {\bf A 15}\ (2000)\ 1749.

\bibitem{bound}
A.~Garc\'{\i}a, R.~Huerta, and G.~S\'anchez-Col\'on,
Phys.\ Lett.\ {\bf B 498}\ (2001)\ 251.

\bibitem{gibson}
W.~M.~Gibson and B.~R.~Pollard,
{\it Symmetry Principles in Elementary Particle Physics}
(Cambridge: Cambridge University Press 1976).

\bibitem{rmf45}
A.~Garc\'{\i}a, R.~Huerta, and G.~S\'anchez-Col\'on,
Rev.\ Mex.\ Fis.\ {\bf 45}\ (1999)\ 244.

\bibitem{limit}
A.~Garc\'{\i}a, R.~Huerta, and G.~S\'anchez-Col\'on,
J.\ Phys.\ G:\ Nucl.\ Part.\ Phys.\ {\bf 26}\ (2000)\ 1417.

\bibitem{hara}
Y. Hara,
Phys.\ Rev.\ Lett.\ {\bf 12}\ (1964)\ 378.

\bibitem{bourquin}
M.~H. Bourquin {\it et al.,\ }Nucl.\ Phys.\ {\bf B 241} (1984) 1.

\bibitem{chan}
A.~W.~Chan {\it et al.,\ }Phys.\ Rev.\ {\bf D58} (1988) 072002.

\bibitem{luk}
B.~Luk {\it et al.,\ }Phys.\ Rev.\ {\bf D38} (1988) 19.

\bibitem{bellefon}
A.~de~Bellefon {\it et al.,\ }Nuovo\ Cim.\ {\bf 28A} (1975) 289.

\bibitem{ross}
R.~T.~Ross {\it et al.,\ }Purdue.\ Conf.\ (1973) 355.

\bibitem{baltay}
C.~Baltay {\it et al.,\ }Phys.\ Lett.\ {\bf 42B} (1972) 129.

\bibitem{kirsch}
L.~E.~Kirsch {\it et al.,\ }Phys.\ Rev.\ {\bf D5} (1972) 1559.

\bibitem{lopez}
G.~Lopez Castro and A. Mariano, Nucl.\ Phys.\ {\bf A 697} (2002) 440.

\end{references}
\end{document}